\documentclass{ws-p8-50x6-00}
\def\ifmath#1{\relax\ifmmode #1\else $#1$\fi}%
\def\rmt{\ifmath{{\mathrm{t}}}} \def\rmcut{\ifmath{{\mathrm{cut}}}}
\newcommand{\QCD}{{\sc qcd}} \newcommand{\NFM}{{\sc nfm}}

    \def\fsz{\footnotesize}
\def\pt{{p_{\rmt}}} \def\vf{\varphi} \def\yct{y_{\rmcut}} \def\kt{k_{\rmt}}
  \def\EE{e$^+$e$^-$}  \def\cl{\centerline}
\def\beqar{\begin{eqnarray}} \def\eeqar{\end{eqnarray}} \def\vs{\vskip}

\begin{document}

\title{On the Scale of Visible Jets in High Energy
electron-positron Collisions}

\author{Liu Lianshou, \  Chen Gang \ and \ Fu Jinghua}

\address{Institute of Particle Physics, Huazhong Normal University,
Wuhan 430079 China}

\author{Presented by \ Liu Lianshou}

\address{E-mail: liuls@iopp.ccnu.edu.cn}

\maketitle

\abstracts{
A study of the dynamical fluctuation property of jets is carried out
using the Monte Carlo method. The results suggest that, 
the anisotropy of dynamical fluctuations in the hadronic system inside jets,
changes abruptly with the variation of the cut parameter $\yct$.
A transition point exists, where these fluctuations 
behave like those in soft hadronic collisions,
i.e. being circular in the transverse plane with respect to dynamical
fluctuations.}

The presently most promising theory of strong interaction
--- Quantum Chromo-Dynamics (\QCD) has the special property of both asymptotic
freedom and colour confinement. For this reason, in any process, even though
the energy scale, $Q^2$, is large enough for perturbative \QCD\ (p\QCD)
to be applicable, there must be a non-perturbative hadronization phase
before the final state particles can be observed. Therefore, the transition
or interplay between hard and soft processes is a very important problem.

An ideal ``laboratory'' for studying this problem is hadron production
in moderate-energy \EE collisions, e.g. at c.m. energy 
about in the range [10, 100] GeV. The initial condition in these processes
is simple and clear. It can safely be considered as a quark-antiquark pair,
moving back to back with high momenta. On the contrary, in other processes,
e.g. in hadron-hadron collisions, the initial condition is complicated with
the problem of hadron structure involved. 

Theoretically, the transition between perturbative and non-perturbative
\QCD\ is at a scale $Q_0 \sim$ 1--2 GeV. Experimentally, the transition
between hard and soft processes is determined by the identification of
jets through some jet-finding process, e.g. the 
Durham algorithm. In these processes, there is a parameter
--- $\yct$, which, in the case of the Durham algorithm,
is essentially the relative transverse momentum $\kt$
squared,  $\kt = \sqrt{\yct} \cdot \sqrt s$.
From the experimental point of view, $\kt$ can be taken as the transition
scale between hard and soft. Its value depends on the definition of ``jet''.

Historically, the discovery in 1975~\cite{Hanson} of a two-jet structure
in \EE\ annihilation at c.m. energies $\geq 6$ GeV has been taken
as an experimental confirmation of the parton model\cite{JELIS}, and the
observation in 1979 of a third jet in e$^+$e$^-$ collisions at 17 -- 30 GeV
has been recognised as the first experimental evidence of the
gluon~\cite{Brandelik}.
These jets, being directly observable in experiments
as ``jets of particles'', will be called ``visible jets''. 
Our aim is to find the scale
corresponding to these visible jets and to discuss its meaning.

For this purpose, let us recall that
the qualitative difference between the typically soft process ---  moderate
energy hadron-hadron collision --- and the typically hard process --- high
energy e$^+$e$^-$ collision --- can be observed most clearly in the property
of dynamical fluctuations therein~\cite{FFLPRD}.  
The latter can be characterized as usually by the anomalous
scaling of normalized factorial moments (\NFM)~\cite{BP}:
\beqar   %%% (1)
  F_q(M)&=&{\frac {1}{M}}\sum\limits_{m=1}^{M}{{\langle n_m(n_m-1)
     \cdots (n_m-q+1)\rangle }\over {{\langle n_m \rangle}^q}}\\ \nonumber
   &\propto& (M)^{\phi_q}\ \  \quad \quad (M\to \infty) \ \ ,
\eeqar
where a region $\Delta$ in 1-, 2- or 3-dimensional phase space is
divided into $M$ cells, $n_m$  is the multiplicity in the $m$th cell,
and $\langle\cdots\rangle$ denotes vertically averaging over the event
sample.
Note that when the fluctuations exist in higher-dimensional (2-D
or 3-D) space, the projection effect~\cite{Ochs} 
will cause the
second-order 1-D \NFM\ to go
to saturation according to the rule\footnote{In order to eliminate the
influence of momentum conservation~\cite{MMCN}, the first few points
($M=1,2$ or 3) should be omitted when fitting the data to Eq.(2).}:
\be   %%% (2)
 F_2^{(a)}(M_a) = A_a-B_a M_a^{-\gamma_a}, \ \
\ee
where $a=1,2,3$ denotes the different 1-D variables. The parameter $\gamma_a$
describes the rate of approach to saturation of the \NFM\
in direction $a$ and is the most
important characteristic for the higher-dimensional dynamical fluctuations.
If $\gamma_a = \gamma_b$, the fluctuations are isotropic in the $a,b$
plane. If $\gamma_a \neq \gamma_b$, the fluctuations are anisotropic
in this plane. The degree of anisotropy is characterized by the Hurst
exponent $H_{ab}$, which can be obtained from the values of $\gamma_a$
and $\gamma_b$ as~\cite{ZGKX} $ H_{ab} = {(1+\gamma_b) / (1+\gamma_a)}. $
The dynamical fluctuations are isotropic when  $H_{ab} = 1$, and anisotropic
when $H_{ab} \neq 1$.

For the 250 GeV/$c$ $\pi$(K)-p collisions from NA22, the Hurst
exponents are found to be~\cite{NA22}: $ H_{\pt\vf}=0.99 \pm 0.01$, 
$ H_{y\pt}=0.48 \pm 0.06$, $H_{y\vf}=0.47 \pm 0.06, $
which means that the dynamical fluctuations in this moderate-energy
hadron-hadron collisions are isotropic in the
transverse plane and anisotropic in the \ longitudinal-transverse planes.
\ This is what should be expected~\cite{WLprl}, \ \ because 

\begin{center}
\begin{picture}(250,450)
\put(-50,329)
{
{\epsfig{file=fig1.epsi,width=340pt,height=120pt}}
}   
\end{picture}
\end{center}

\vskip-11.3cm
\cl{Fig.1~The variation of the parameter $\gamma$ with $\yct$ ($\kt$)} 

\vskip0.5cm

\noindent 
there are 
almost no hard collisions at this energy and the direction of motion of 
the incident hadrons (longitudinal direction) should be previleged. 

In high energy e$^+$e$^-$ 
collisions, the longitudinal direction is chosen
along the thrust axis, which is the direction of motion of the primary
quark-antiquark pair. Since this pair of quark and antiquark moves back to 
back with very high momenta, the magnitude of the average momentum of 
final state hadrons is also anisotropic due to momentum conservation.
However, the dynamical fluctuations in this case come from
the \QCD\ branching of partons~\cite{Vineziano}, which is isotropic
in nature. Therefore, in this case
the dynamical fluctuations should be
isotropic in 3-D phase space. 
A Monte Carlo study for e$^+$e$^-$ collisions at 91.2 GeV confirms this
assertion~\cite{FFLPRD}. 
Also
the presently available experimental data on e$^+$e$^-$ collisions at 91.2 GeV
show isotropic dynamical fluctuations in 3-D~\cite{DELPHI}.
\noindent 

Now we apply this technique to the ``2-jet'' sub-sample of \EE collisions
obtained from a
certain, e.g. Durham, jet-algorithm with some definite value of $\yct$.
Doing the analysis for different values of $\yct$,
the dependence of dynamical-fluctuation property of the ``2-jet''
sample on the value of $\yct$ can be investigated.
Two event samples are constructed from the Jetset7.4 and Herwig5.9 generators,
each consisting of 400 000 \EE\ collision events at c.m.
energy 91.2 GeV.  The 
variation of $\gamma$'s of the 2-jet sample
with $\yct$ ($\kt$) are shown in Fig's 1($a$) and ($b$), respectively.
It shows an interesting pattern. When
$\yct$ ($\kt$) is very small, the three  $\gamma$'s are separate. As
$\yct$ ($\kt$) increases, $\gamma_\pt$ and $\gamma_\vf$ approach each
other and cross over sharply at a certain point. After that, the
three $\gamma$'s approach a common value. The latter is
due to the fact that when
$\yct$ is very large, the ``2-jet'' sample coincides with the full sample
and the dynamical fluctuations in the latter are isotropic.  

We will call the point where $\gamma_\pt$ crosses $\gamma_\vf$
the {\em transition point}. It has the unique property $\gamma_\pt =
\gamma_\vf \neq \gamma_y$, i.e., the jets at this point are circular in
the transverse plane with respect to dynamical fluctuations.
These jets will, therefore,  be called {\em circular jets}.

The above-mentioned results are qualitatively the same for the two
event generators, but the $\yct$ ($\kt$) values
at the transition point are somewhat different.
The cut parameters $\yct$, the values of the $\gamma$, the corresponding 
Hurst exponents $H$ and the relative transverse momenta $\kt$ at the
transition point are listed in Table I.

\vskip0.3cm
\noindent
\cl{Table I \ $\gamma$, $H$, 
$\yct$ (GeV/$c$) and $\kt$ (GeV/$c$) at the transition point}
\vskip-0.8cm

\def\bcc{\begin{center}} \def\ecc{\end{center}}
\def\btbl{\begin{tabular}} \def\etbl{\end{tabular}}
\def\fsz{\footnotesize}
\bcc
\btbl{|c|c|c|c|c|c|c|c|c|c|}\hline
& $\yct$ & $\gamma_y$ & $\gamma_\pt$ & $\gamma_\vf$ &
$H_{y\pt}$ & $H_{y\vf}$ & $H_{\pt\vf}$ & $\kt$ \\ \hline
\fsz Jetset & \fsz 0.0048 & \fsz 1.074 & \fsz 0.514 & \fsz 0.461 & \fsz 0.73 
& \fsz 0.70 & \fsz  0.96 &  \fsz 6.32 \\
\fsz 7.4 & {\fsz $\pm$0.0007}  & {\fsz $\pm$0.037} & {\fsz $\pm$0.080}
     & {\fsz $\pm$0.021} & {\fsz $\pm$0.06}
     & {\fsz $\pm$0.06} & {\fsz $\pm$0.10} & {\fsz $\pm$0.03} \\ \hline
\fsz Herwig & \fsz 0.0022 & \fsz 1.237 & \fsz 0.633 & \fsz 0.637 & \fsz 0.73 
& \fsz 0.73 &  \fsz 1.00 & \fsz 4.28 \\
\fsz 5.9 & {\fsz $\pm$0.0008}  & {\fsz $\pm$0.066} & {\fsz $\pm$0.064}
     & {\fsz $\pm$0.051} & {\fsz $\pm$0.05}
     & {\fsz $\pm$0.05} & {\fsz $\pm$0.07} & {\fsz $\pm$0.02} \\ \hline
\etbl
\ecc

\vskip0.5cm
It is natural to ask the question: Is there any relation between the
{\em circular jets} determined by the condition $\gamma_\pt = \gamma_\vf
\neq \gamma_y$ and the {\em visible jets} directly observable in experiments
as ``jets of particles''?  
In order to answer this question, we
plot in Fig. 2 the ratios $R_2$ and $R_3$ of
``2-jet'' and ``3-jet'' events as functions of the  relative transverse
momentum $\kt$ at different c.m. energies.

Let us consider the point where a third jet starts to appear.
Historically, a third jet was firstly observed in \EE\ collisions
at c.m. energy 17 GeV. It can be seen from Fig.2 that, for $\sqrt s=$ 17 GeV,
$R_3$ starts to appear at around $\kt=$ 8--10 GeV/$c$, cf. the dashed vertical
lines in Fig. 2. This value of $\kt$ is consistent with
the $\kt$ value (4.3--6.3 GeV/$c$) of a circular jet within a factor of 2,
cf. Table I.
Thus we see that the {\em circular jet}, defined as a kind of jet
circular in the transverse plane with respect to dynamical fluctuations,
and the {\em visible jet}, defined as a kind of jet directly
observable in experiments as a ``jet of particles'', have about the same
scale --- $\kt\sim$ 5--10 GeV/$c$.

\vs0.5cm
\cl{Table II The values of $\yct$ and $\kt$ at the transition points}  
\def\bcc{\begin{center}} \def\ecc{\end{center}}
\def\btbl{\begin{tabular}} \def\etbl{\end{tabular}}
\bcc\btbl{|c|c|c|c|c|c|}\hline
$\sqrt s$ {\fsz (GeV)} & $\yct$ & $\kt$ {\fsz (GeV/$c$)} \\ \hline
\fsz 50  & \fsz 0.0186 {\fsz $\pm$0.0012} 
& \fsz 6.82 {\fsz $\pm$0.03} \\ \hline
\fsz 30  & \fsz 0.059 {\fsz $\pm$0.002} 
& \fsz 7.28 {\fsz $\pm$0.03} \\ \hline
\etbl\ecc
\vskip0.5cm

In order to check how sensitively the magnitude of
this scale depends on the c.m. energy of \EE collisions, a similar analysis
is caried out for $\sqrt s=$ 50 and 30 GeV using \ Jetset7.4, \ cf.
Fig's.1 ($c$, $d$). \ It can be seen that although 

\vskip0.3cm

\begin{center}
\begin{picture}(250,450)
\put(-42,265)        
{
{\epsfig{file=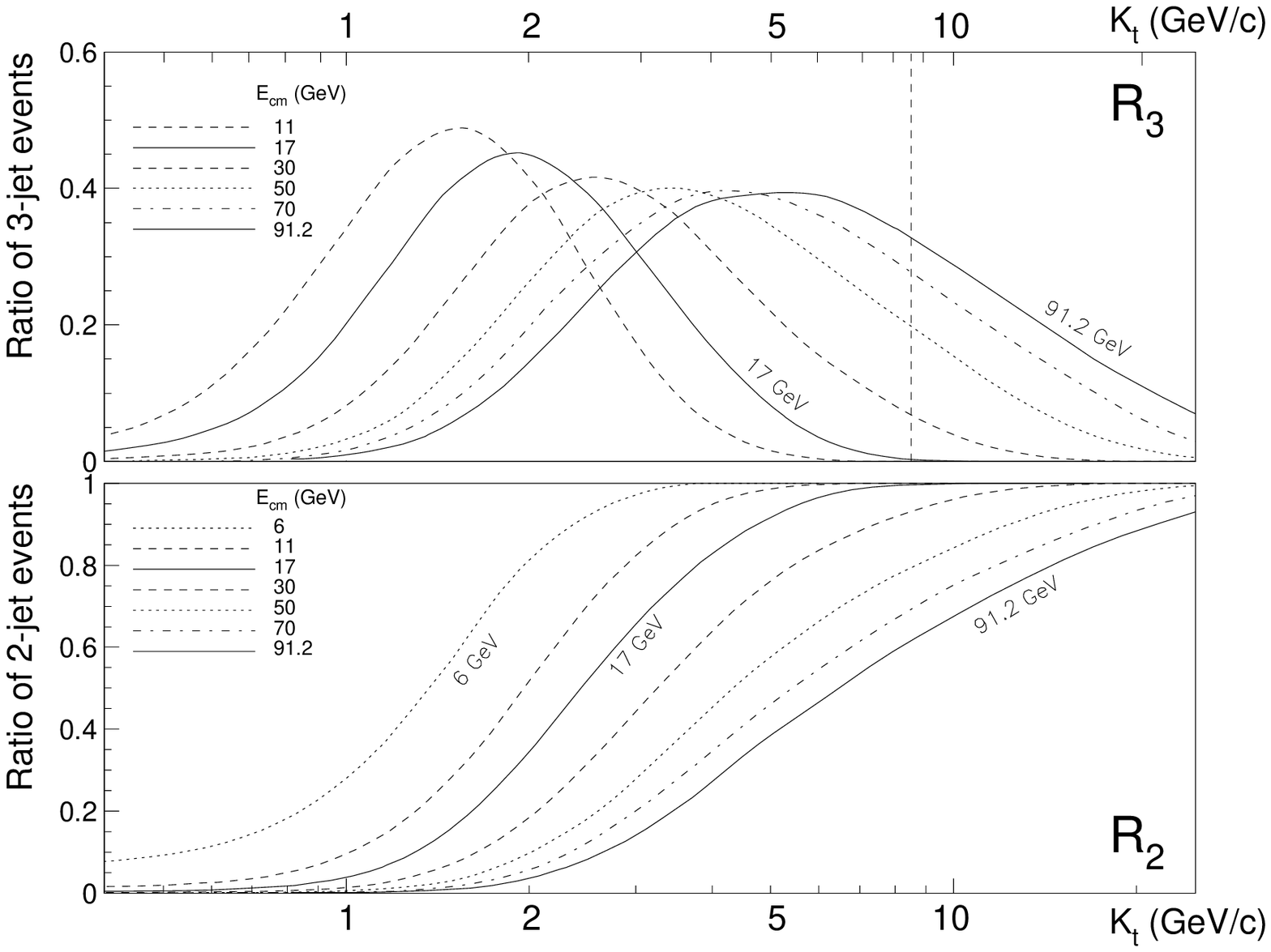,width=180.0pt,height=210pt}}
}    
\put(107,265)
{
{\epsfig{file=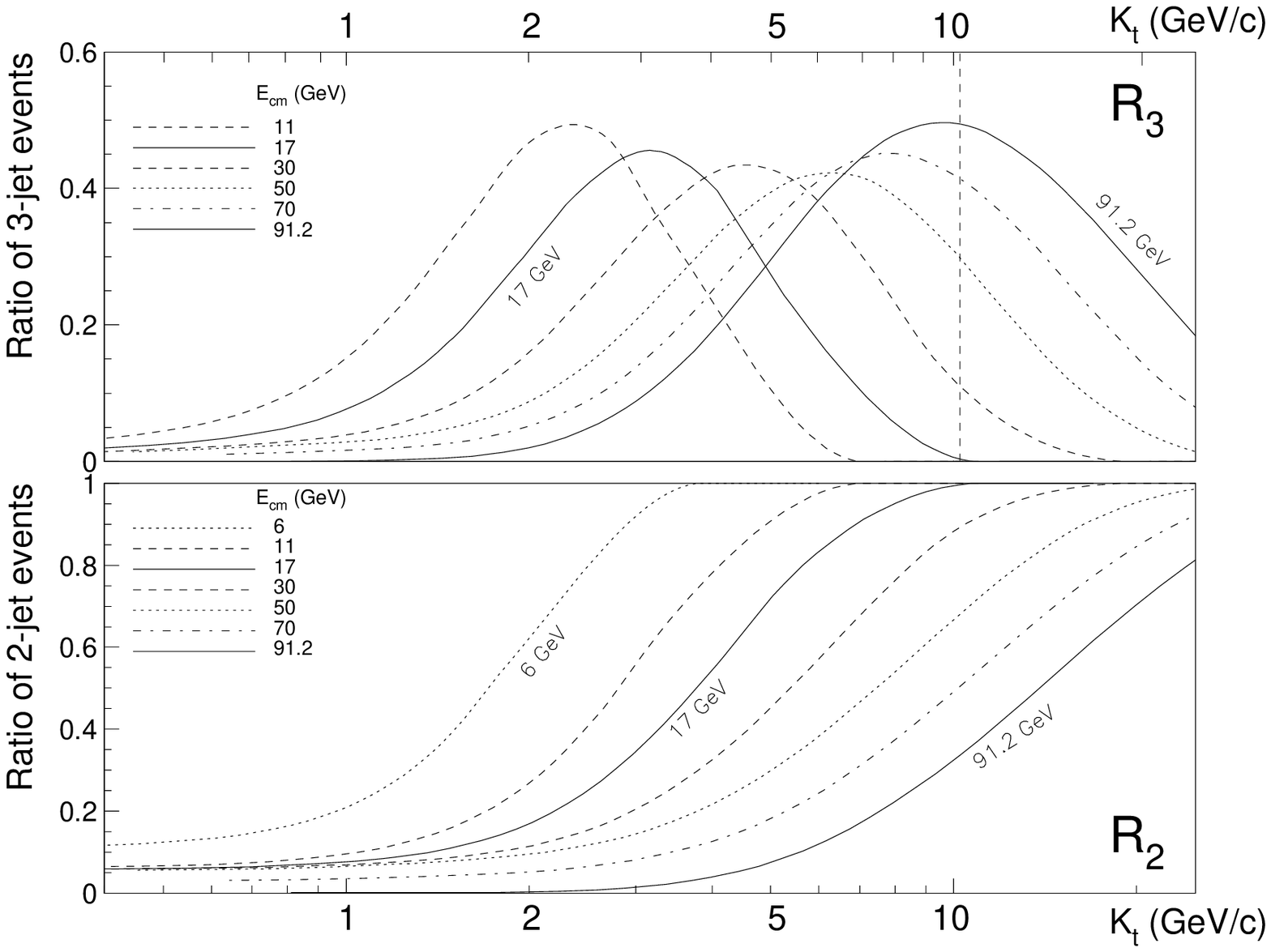,width=180.0pt,height=210pt}}
}
\end{picture}
\end{center}

\vskip-13.2cm
\hskip0.8cm ($a$) \hskip4.7cm ($b$)
 \vskip1.0cm

\hskip0.6cm{\small Fig.2 \ The ratios $R_3$ and $R_2$ of 3- and 2-jet events as
functions of $\kt$

\hskip1.2cm at different c.m. energies, ($a$) from Jetset7.4;
($b$) from Herwig5.9}

\vskip0.2cm

\noindent 
the shape of
$\gamma_i$ versus $\yct$ ($\kt$) ($i=y,\pt,\vf$) changes considerably with
energy the qualitative trend is the same for these energies. In particular,
the transition point where $\gamma_\pt$ crosses $\gamma_\vf$ exists in
all cases. The values of $\yct$ and $\kt$ at the 
transition point are listed in Table II. It can be seen that the $\kt$ 
values are also in the range 5--10 GeV/$c$. This shows that the sacle 
$\kt\sim$ 5--10 GeV/$c$ for 
the {\em circular jet} is universal, at least
for moderate energy \EE collisions.

This scale is to be compared with the scale $\kt\sim$ 1--2 GeV/$c$,
which is the scale for the transition between the perturbative and
non-perturbative domains. It is interesting also to see what happens in
the results of jet-algorithm at this scale.
It can be seen from Fig.2$a$ (Jetset7.4) that, at this scale
($\kt\sim$ 1--2 GeV/$c$)
the ratio $R_2$ of ``2-jet'' events tends to vanish
almost independently of energy, provided the latter is not too low.
This can be explained as follows.
Consider, for example, an event with only two hard partons, having no
perturbative branching at all.  Even in this case, the two
partons will still undergo non-perturbative hadronization to produce
final-state particles. If the $\kt$ is chosen to be less than 1--2 GeV/$c$,
then the non-perturbative hadronization with small transverse momentum will
also be considered as the production of new ``jets'' and this
``should-be'' 2-jet event will be taken as a ``multi-jet'' 
event too. This means that, when
$\kt <$ 1--2 GeV/$c$, events with small transverse momentum will also
become ``multi-jet'' ones, and $R_2$ vanishes.
However, even when $\kt <$ 1--2 GeV/$c$,
a few 2-jet events may still survive if the hadronization
is alomst colinear. This effect becomes observable
when the energy is very low, see, e.g., the
$R_2$ curve for $\sqrt s=6$ GeV in Fig.2$a$.
A similar picture holds also for the results from Herwig5.9, cf. Fig.2$b$,
but the almost-colinear hadronization appears earlier.

Let us give some comments on the physical picture
behind the above-mentioned two scales. A circular (or visible) jet is
originated from a hard parton. The production of this parton is a
hard process. Its evolution
into final state particles includes a perturbative branching and subsequent
hadronization. The hadronization is a soft process. The perturbative
branching (sometimes called parton shower) between the hard production and
soft hadronization connects these two processes. This perturbative branching
inside a circular jet
is certainly not soft, but is also not so hard. This kind of processes
is sometimes given the name {\em semi-hard} in the literature.
The isotropic property of dynamical fluctuations provides a criterion for
the discrimination of the hard production of circular jets and the
(semi-hard) parton shower inside these jets.

\section*{Acknowledgments}
Supported in part by the National Natural Science Foundation of China
(NSFC) under Project 19975021.  The authors are grateful to Wolfram Kittel, 
Wu Yuanfang and Xie Qubin for valuable discussions.

\end{document}

%%%%%%%%%%%%%%%%%%%%%%%%%%%%%%%%%%%%%%%%%%%%%%%%%%%%%%%%%%%%%%%%%%%%%%%%%%%%%
%% End of  ws-p8-50x6-00.tex  
%%%%%%%%%%%%%%%%%%%%%%%%%%%%%%%%%%%%%%%%%%%%%%%%%%%%%%%%%%%%%%%%%%%%%%%%%%%%%